\documentclass{webofc}
\usepackage[varg]{txfonts}   
\usepackage{subcaption}

\usepackage[dvipsnames]{xcolor}
\usepackage[colorlinks=true,linkcolor=magenta,urlcolor=blue,citecolor=blue]{hyperref}
\usepackage{cleveref}
\usepackage{fancyhdr}  % Package for headers and footers
\begin{document}
 \title{Electroweak phase transition in a vector dark matter \\ scenario}

\author{\firstname{Nico} \lastname{Benincasa}\inst{1}\fnsep\thanks{\email{nico.alexis.benincasa@ut.ee}} 
\and
\firstname{Luigi} \lastname{Delle Rose}\inst{2,3}\fnsep\thanks{\email{luigi.dellerose@unical.it}} \and
\firstname{Luca} \lastname{Panizzi}\inst{2,3}\fnsep\thanks{\email{luca.panizzi@unical.it}}\and
\firstname{Maimoona} \lastname{Razzaq}\inst{2,3}\fnsep \thanks{poster author, \email{maimoona.razzaq@unical.it}}
  \and
\firstname{Savio} \lastname{Urzetta}\inst{2}\fnsep\thanks{\email{sav.urze98@gmail.com}} 
}

\institute{School of Physics, University of Electronic Science and Technology of China, 611731 Chengdu, China    
\and
Dipartimento di Fisica, Università della Calabria, Arcavacata di Rende, I-87036, Cosenza, Italy
\and
    INFN, Gruppo Collegato di Cosenza, Arcavacata di Rende, I-87036, Cosenza, Italy        
}

\abstract{%
   This study explores the parameter space of a minimal extension of the Standard Model with a non-abelian $SU(2)$ group, in which the gauge bosons are stable and acquire mass through a mechanism of spontaneous symmetry breaking involving a new scalar doublet which interacts with the Higgs boson through a quartic coupling. The exploration aims to assess whether it is possible to obtain a first-order phase transition while ensuring that the gauge bosons are viable dark matter candidates. Theoretical, astrophysical and collider bounds are considered. The results are then tested against the sensitivity of future experiments for the detection of gravitational wave signals. 
 }
\maketitle
\section{Introduction}
\label{intro}
Cosmological and astrophysical observations strongly indicate that 
baryonic matter is not the primary component of matter in the Universe \cite{bertone2005particle}. Non-baryonic matter manifests through gravitational effects, but it does not interact with Standard Model (SM) particles via the electromagnetic or quantum chromodynamics forces, and is therefore referred to as dark matter (DM). Many theories have been put forward to extend the SM with new DM particles. If a DM state interacts with SM fields (weakly, as it has so far escaped detection), it is in principle possible to pose bounds on its mass and on the size of the interactions.
DM models can be constrained via direct and indirect detection experiments, and at particle accelerators such as the Large Hadron Collider. The relative importance of such constraints depends on the hypotheses about the interactions of DM with SM particles. Models where the DM interacts exclusively through the Higgs boson are in general less constrained due to the uncertainties in the measurement of Higgs couplings. 
Since recently, gravitational waves (GWs) offer a novel and complementary approach to exploring these models, especially those that involve additional scalar fields capable of triggering first-order phase transitions, thus leading to GW emissions. The main focus of our work is to explore one of such models, where the DM is represented by the gauge bosons of a new non-abelian gauge group, which acquire mass through a mechanism of spontaneous symmetry breaking in the dark sector driven by a new scalar field.
 
\section{The Model}
\label{sec-1} 
We explore an extension of the SM that includes a new $SU(2)_D$ gauge symmetry, under which all SM particles are singlets. In this framework, we introduce a scalar doublet that breaks the $SU(2)_D$ symmetry in the dark sector through a Higgs mechanism. The custodial $SO(3)$ symmetry left in the dark sector after spontaneous symmetry breaking ensures that three gauge bosons remain stable and mass-degenerate, with a common mass of $m_{V_{D}} ={g_D v_D}/{2}$, where $g_D$ is the $SU(2)_D$ gauge coupling and $v_D$ is the vacuum expectation value of the dark scalar doublet. Because the gauge bosons are triplet under $SO(3)$ and all other particles are singlets, this custodial symmetry prohibits the decay of the these gauge bosons \cite{Hambye:2008bq}.
The Lagrangian of the model is given by
\begin{equation}
\mathcal{L} = \mathcal{L}_{SM} - \frac{1}{4} F_{D\mu\nu} F^{\mu\nu}_D + (D_\mu \Phi_D)^\dagger (D^\mu \Phi_D) + \mu_D^2 (\Phi_D^\dagger \Phi_D) - \lambda_D (\Phi_D^\dagger \Phi_D)^2 - \lambda_{HD} (\Phi_D^\dagger \Phi_D)(\Phi^\dagger \Phi),
\end{equation}
where $\mathcal{L}_{SM}$ is the SM Lagrangian, $F_D^{\mu\nu}$ is the field strength associated to $SU(2)_D$, $\Phi_D$ is the dark scalar doublet, and $\Phi$ is the SM Higgs doublet (see also \cite{Belyaev:2022shr,EWPTSU2Dark}). After the Electroweak and dark symmetry breakings, we can write the scalar doublets in the unitary gauge as
\[
\Phi = \frac{1}{\sqrt{2}} \begin{pmatrix} 0 \\ v + h_1 \end{pmatrix}, \quad \Phi_D = \frac{1}{\sqrt{2}} \begin{pmatrix} 0 \\ v_D + h_2 \end{pmatrix},
\]
where $v$ and $v_D$ are the vacuum expectation values of the Electroweak and SU(2)$_D$ symmetries, respectively, and $h_1$, $h_2$ are real scalar fields in the interaction eigenbasis. The portal coupling $\lambda_{HD}$ between the SM and the dark sector induces a mixing between the two scalar fields $h_1$ and $h_2$. The free parameters of the model are four, and they can be chosen to be the gauge coupling $g_D$, the gauge boson mass $m_{V_D}$, the mass of the new scalar $m_{H_D}$ and the mixing angle $\theta_D$ between the two real scalars in the mass eigenbasis, $h$ and $H_D$.

\section{Constraints}
\label{sec-3}
The parameter space of the model is subject to several constraints, ensuring theoretical consistency and agreement with experimental data:
\begin{itemize}
    \item \textbf{Theoretical Constraints:} Perturbative unitarity and boundedness from below are applied to ensure the theoretical consistency of the model.
    \item \textbf{Collider Constraints:} Bounds are imposed based on direct searches for new scalar particles and on the compatibility with the measured Higgs couplings at the LHC \cite{bahl2023higgstools}.
    \item \textbf{Electroweak Precision Tests:} Electroweak precision observables S, T and U \cite{Peskin:1992sw,Peskin:1990zt} are used to parameterize the effects of new physics on Electroweak observables.
    \item \textbf{Dark Matter Relic Density:} The relic density of DM, as measured by Planck \cite{aghanim2020planck},
    imposes stringent constraints on the parameters. The DM relic density, obtained through the freeze-out mechanism, can be approximated by:
    \[
    \Omega_{DM} \approx 0.1 \frac{x_{f}}{\sqrt{g_*(m_{V_{D}})}}\frac{10^{-8}~{\rm GeV}^{-2}}{\langle \sigma v \rangle}\approx 0.12
    \]
    where $x_{f}$ is the freeze-out parameter, $g_*(m_{V_D})$ represents the effective number of relativistic degrees of freedom at temperature $T=m_{V_D}$ and $\langle \sigma v \rangle$ is the thermally averaged annihilation cross-section. There are several annihilation channels at tree-level that influence the relic density of vector DM during  freeze-out, shown in \cref{fig:annihilationtopologies}. Notably, the annihilation channel on the right exhibits a unique characteristic, as it involves one DM candidate in the final state, a process that is not possible in models based on $Z_2$ symmetry \cite{Hambye:2008bq}.
\begin{figure}[h]
    \centering
    
    \begin{subfigure}[]{0.3\textwidth}
        \centering
        \includegraphics[width=\textwidth]{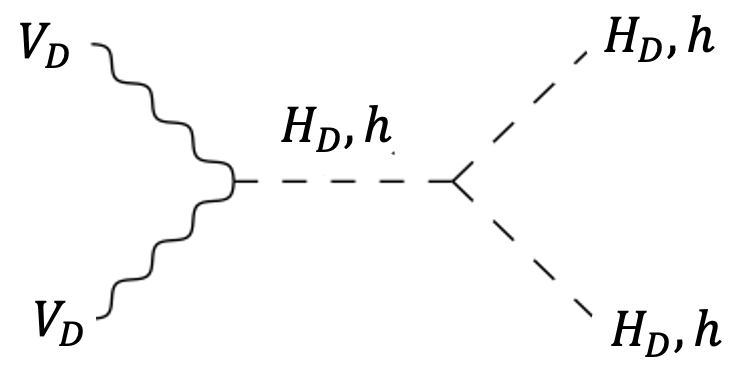}
        %\caption{}
        \label{fig:sub1}
    \end{subfigure}
    \hfill
    \begin{subfigure}[]{0.2\textwidth}
        \centering
        \includegraphics[width=\textwidth]{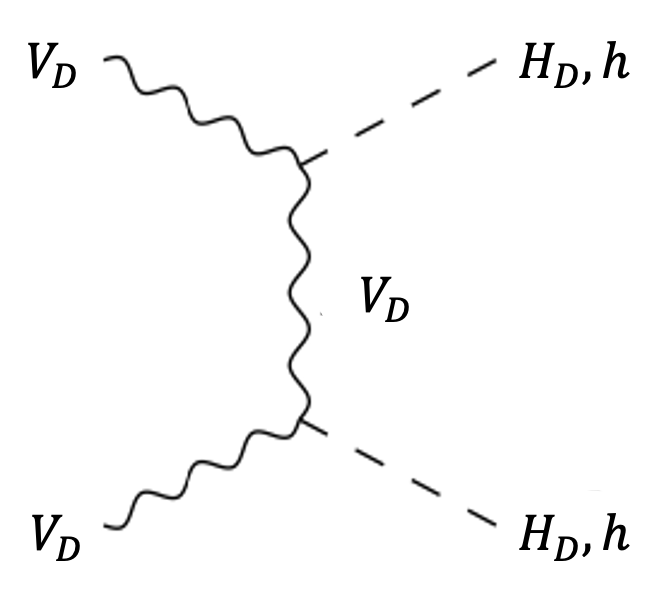}        
        %\caption{}
        \label{fig:sub2}
    \end{subfigure}
    \hfill
    \begin{subfigure}[]{0.3\textwidth}
        \centering
        \includegraphics[width=\textwidth]{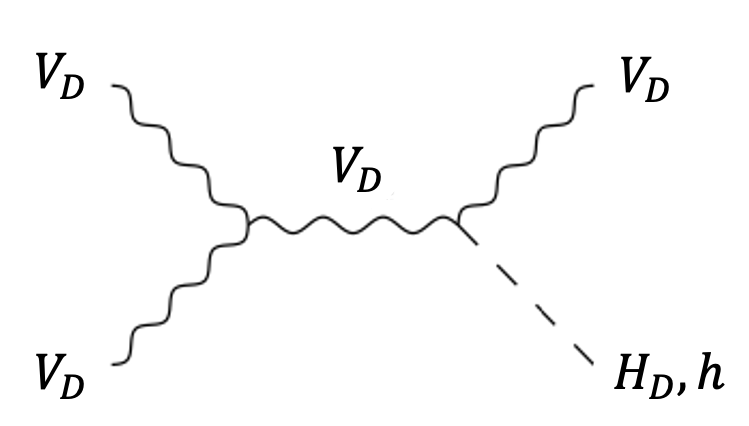}
        %\caption{}
        \label{fig:sub3}
    \end{subfigure}    
    \caption{Topologies describing DM annihilation processes.}
    \label{fig:annihilationtopologies}
\end{figure}
    \item \textbf{Direct and Indirect Detection:} Constraints from direct detection experiments (e.g., XENON1T, LUX-Zepelin) \cite{aprile2017first, aprile2018dark, collaboration2020lux} and indirect detection experiments (e.g., HESS, FERMI LAT) \cite{Fermi-LAT:2015att,abdalla2022search} are applied to further narrow down the parameter space.    
\end{itemize}

\section{Gravitational Waves from First-Order Phase Transitions}
\label{sec-2}
In a first-order phase transition, the system tunnels from a metastable to the ground state through bubble nucleation. In the early universe the bubbles expand, collide and eventually convert energy into a background of GWs. The power spectrum of these can be entirely characterized by the following phase transition parameters, which can be, in turn, computed from the Lagrangian of the model:
\begin{itemize}
    \item $\alpha$: the strength of the phase transition, defined as
    \begin{equation}
         \alpha \equiv\frac{\Delta\epsilon}{\rho_{\text{rad}}}\Big|_{T=T_n},   \quad \Delta \epsilon \equiv \epsilon \big |_{\text{false vacuum}} - \epsilon \big |_{\text{true vacuum}}
    \end{equation}
where $\Delta\epsilon$ is the vacuum energy released during the phase transitions and $\rho_{\text{rad}}=\frac{\pi^2}{30} g_* T^4$ \cite{Espinosa:2010hh} and $T_n$ is the nucleation temperature.
\item $\beta/H$: the inverse of the transition time duration defined as follows \cite{Grojean:2006bp}:
\begin{equation}
    \label{eq:beta}
    \frac{\beta}{H} = T \frac{d(S/T)}{dT} \Big|_{T=T_n}
\end{equation}
where $S$ represents the 3-dimensional Euclidean action. A smaller value of $\beta/H$ typically corresponds to a stronger phase transition, which in turn leads to a more pronounced GW signal.

\item $T_p$: the percolation temperature is the temperature at which bubbles of true vacuum expand enough and merge with one another to form a continuous phase across space such that the false vacuum fraction is $\sim 0.71$. In absence of a significant amount of supercooling, as in our scenario, $T_p \simeq T_n$, where $T_n$ is defined as the temperature at which $\Gamma/H^4 \simeq 1$, with $\Gamma$ the nucleation rate.

\item $v_w$: the bubble wall velocity refers to the speed at which the true vacuum bubble expands through the surrounding plasma, see \cite{DeCurtis:2022hlx, DeCurtis:2023hil, DeCurtis:2024hvh} for a recent discussion. 
 
\end{itemize}

\section{Discussion}
\label{sec-4}

\begin{figure}[h]
\centering
\includegraphics[width=.7\textwidth,clip]{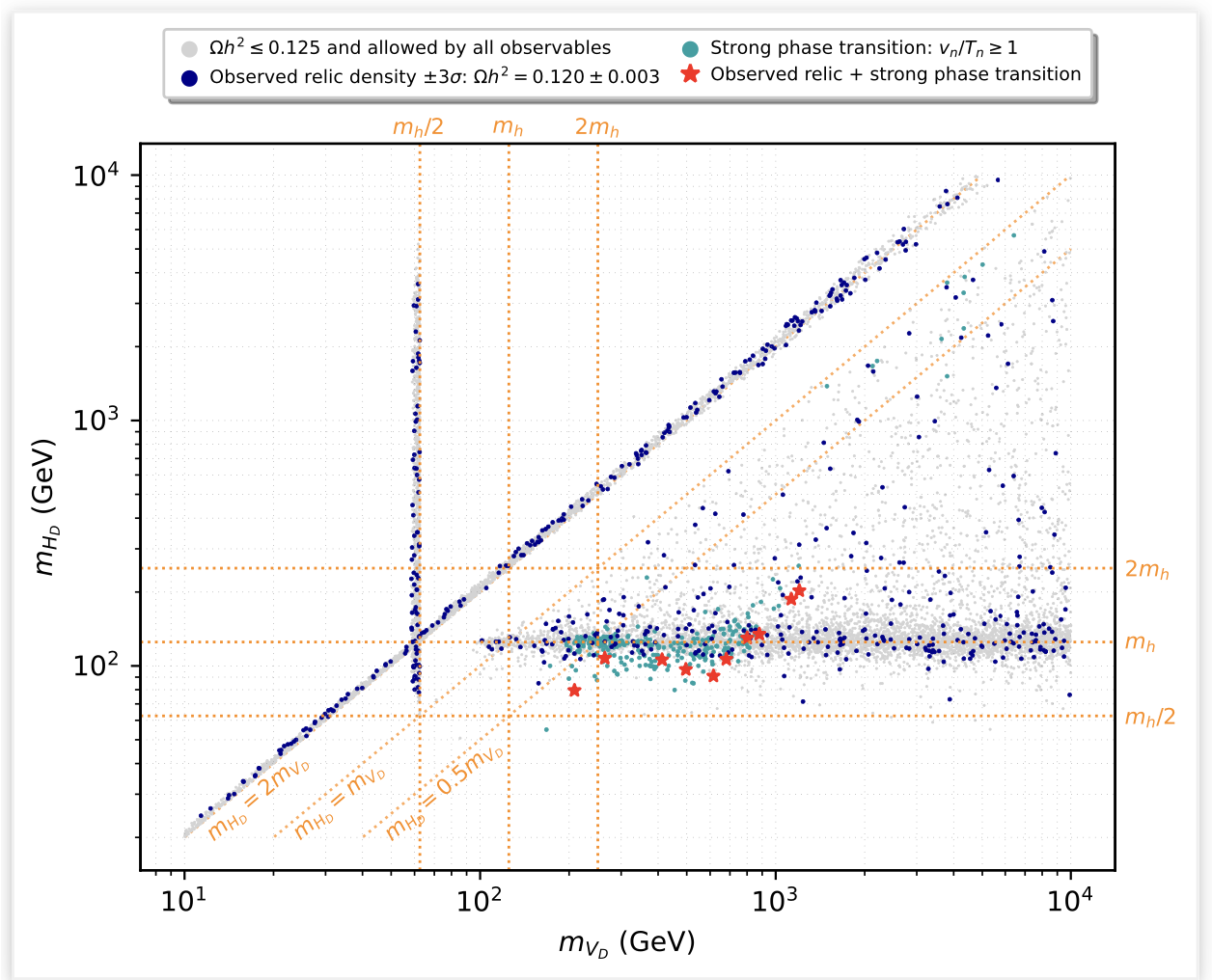}
\caption{Points in the $(m_V,m_{H_D})$ projection of the parameter space which satisfy experimental and theoretical bounds. The gray points corresponds to underabundant scenarios with relic density within $5\sigma$ from the Planck measurement. The blue points reproduce the correct relic density within $3\sigma$. Points in cyan corresponds to a strong first-order phase transition. Red stars correspond to points with a strong phase transition and the correct relic density.}
\label{fig:scatterplot}       
\end{figure}

The parameter space of the model, sifted with both the theoretical and experimental constraints discussed in in \cref{sec-3}, and projected onto the plane $(m_{V_D},m_{H_D})$ is shown in \cref{fig:scatterplot}. The colour code emphasises the interplay between the predicted relic density and the phase transition strength, here quantified by the ratio $v_n/T_n$, with $v_n^2 = (v^2 + v_D^2)|_{T = T_n}$ evaluated at the nucleation temperature. Specifically, the gray points correspond to regions where the relic density is underabundant within 5$\sigma$ of the measured Planck value, {\it i.e.} $\Omega h^2 \leq 0.125$, and all observational constraints (astrophysical and collider) are satisfied. Among such points, those with blue colour indicate regions where the relic density matches the observed value within $3\sigma$, i.e., $\Omega h^2 = 0.120 \pm 0.003$ \cite{aghanim2020planck}, while cyan points mark regions where a strong phase transition occurs, with $v_n/T_n \geq 1$. The red stars highlight points where the relic density is within the observed range and a strong phase transition is present.
In this figure, the presence of two resonant annihilation processes is clearly evident and manifests in the clustering of points along the vertical and oblique directions. When \( m_{H_D} = 2m_{V_D} \), annihilation proceeds through the \( H_D \) resonance, while for \( m_h = 2m_{V_D} \), it proceeds through the \( h \) resonance. The lower triangular region, below \( m_{V_D} = m_{H_D} \) and above \( m_{H_D}= m_h/2 \), is a non-resonant region. The region with a higher density of points, where \( m_{H_D} \approx m_h \) and \( m_{V_D} > m_h \), is less constrained by collider observables.

\begin{figure}[h]
\centering
\includegraphics[width=.7\textwidth]{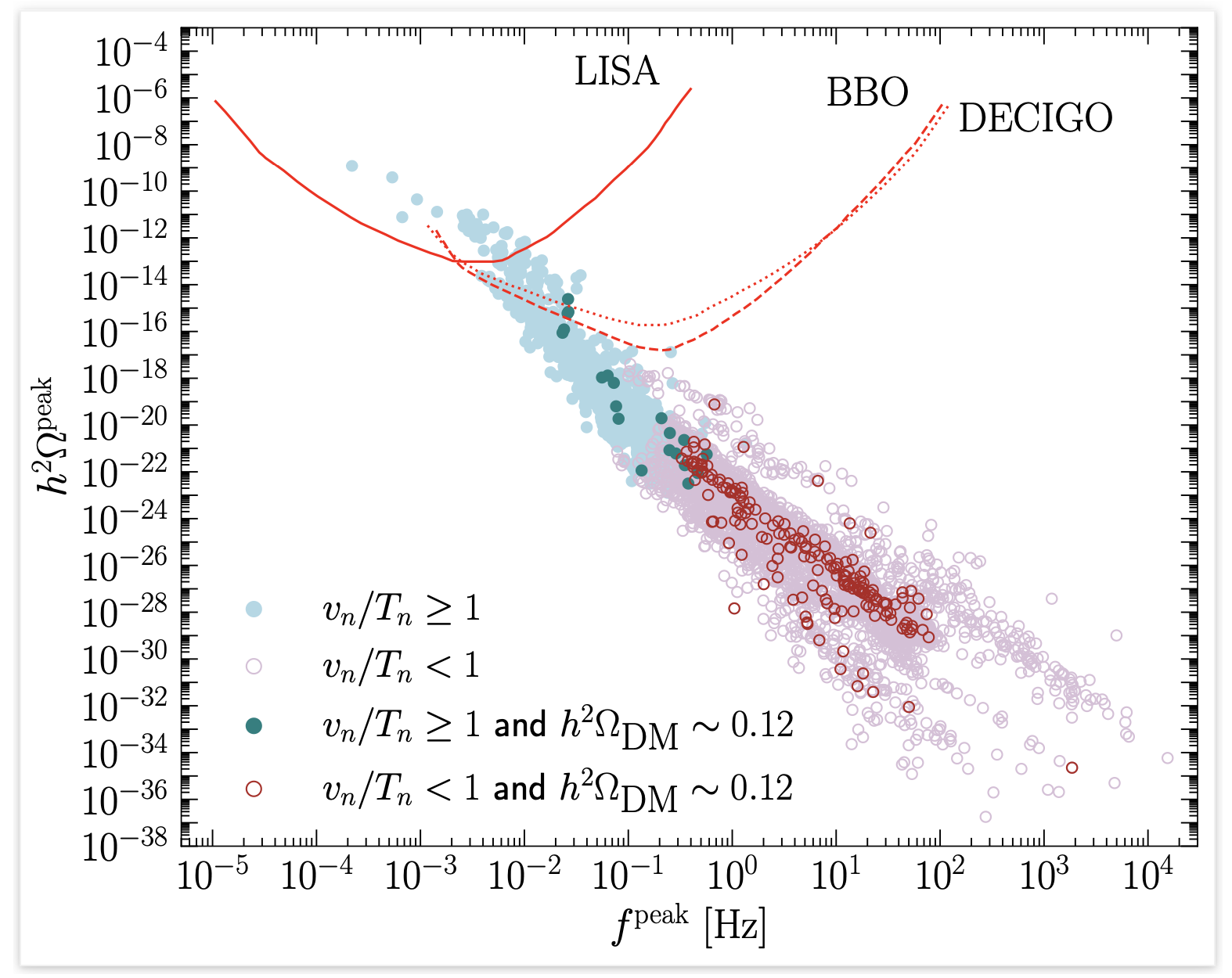}
\caption{GW signal \( h^2 \Omega^{\text{peak}} \) as a function of peak frequency \( f^{\text{peak}} \), with color coding representing the strength of the phase transition and dark matter relic density. Sensitivity curves of future GW detectors are included \cite{LISA:2017pwj,Corbin:2005ny,Seto:2001qf}.}
\label{fig:GW}
\end{figure}
To further explore the connection between DM and GWs, we analyze the GW signal produced during phase transitions. The stochastic background of GWs depends on three main mechanisms: the propagation of sound waves, bubble collisions and turbulence effects in the plasma. Therefore, the GW signal is expressed as \cite{caprini2016science}
\begin{equation}
    h^2 \Omega_{\text{GW}} \simeq h^2 \Omega_{\text{col}} + h^2 \Omega_{\text{sw}} + h^2 \Omega_{\text{turb}}.
\end{equation} 
In \cref{fig:GW}, the GW signal \( h^2 \Omega^{\text{peak}} \) is plotted as a function of the peak frequency \( f^{\text{peak}} \). The color coding in the figure corresponds to the strength of the phase transition and the relic density of DM. Additionally, the sensitivity curves of some future GW detectors are included to show the expected detection capabilities for these signals. 

 \section{Conclusion}
 \label{sec-5}
   We have studied a spin-one DM model based on a $SU(2)_D$ gauge group in a dark sector in which a custodial symmetry ensures the DM stability. The dark gauge bosons acquire masses through the $SU(2)_D$ symmetry breaking triggered by a dark scalar doublet. This symmetry breaking resulted in a phase transition in the dark sector. We analyzed the nature of the phase transition and characterised its property. If a first-order phase transition was found, we examined the corresponding GW signal and assessed its detectability by future GW observatories. While the forthcoming LISA experiment \cite{LISA:2017pwj} could detect GW signals associated with sufficiently strong first-order phase transitions in this model, a much larger sensitivity reach is necessary to probe the region of the parameter space in which the DM candidate could comply with all current constraints while reproducing, at the same time, the correct relic abundance.

\end{document}